\documentclass[english]{article}
\usepackage[T1]{fontenc}
\usepackage[latin9]{inputenc}
\usepackage{geometry}
\usepackage{graphicx}

\makeatletter

\newcommand{\lyxaddress}[1]{
\par {\raggedright #1
\vspace{1.4em}
\noindent\par}
}
\newenvironment{lyxlist}[1]
{\begin{list}{}
{\settowidth{\labelwidth}{#1}
 \setlength{\leftmargin}{\labelwidth}
 \addtolength{\leftmargin}{\labelsep}
 }}
{\end{list}}

\usepackage{babel}
\makeatother

\begin{document}

\title{A Combinatorial Enumeration of Distances for Calculating Energy in
Molecular Conformational Space.}

\author{Jacques Gabarro-Arpa}

\maketitle

\lyxaddress{Ecole Normale Supérieure de Cachan - LBPA CNRS UMR 8113 - 61, Avenue
du Président Wilson - 94235 Cachan cedex - France.}

\lyxaddress{e-mail: jga@gtran.org}

\begin{abstract}
In previous works it was shown that protein 3D-conformations could
be encoded into discrete sequences called dominance partition sequences
(DPS), that generated a linear partition of molecular conformational
space into regions of molecular conformations that have the same DPS.
In this work we describe procedures for building in a cubic lattice
the set of 3D-conformations that are compatible with a given DPS.
Furthermore, this set can be structured as a graph upon which a combinatorial
algorithm can be applied for computing the mean energy of the conformations in a cell.
\end{abstract}

\section{Introduction}

In previous papers {[}1-5] we have built a series of mathematical
tools for studying the multidimensional molecular conformational space
of proteins, with the aim of understanding the dynamical states of
proteins by building a complete energy surface.

In this approach, the 3D-structures of protein molecules are encoded
into a linear sequence of numbers called \textbf{dominance partition sequences}
(DPS) {[}1-5], there are three of these sequences one for each coordinate
$x$, $y$ and $z$. For a molecule of $N$ atoms assigning to each
atom a number in the range $1-N$, then for a coordinate $c$ the
DPS is : \emph{the sequence of atom numbers sorted in ascending order
of the value of the $c$-coordinate of their respective atoms}.

A typical 3D-DP sequence may be something like

$\{\{(5)(3)(1)(4)(2)\}_{x},\{(2)(4)(3)(1)(5)\}_{y},\{(1)(3)(5)(2)(4)\}_{z}\}$

which means that

$x_{5}<x_{3}<x_{1}<x_{4}<x_{2}$, \hspace*{4mm} $y_{2}<y_{4}<y_{3}<y_{1}<y_{5}$,
\hspace*{4mm}$z_{1}<z_{3}<z_{5}<z_{2}<z_{4}$

A simple example of DPS can be extracted from Fig. 1 where the $\alpha$-carbon
skeleton of the pancreatic trypsin inhibitor protein {[}6,7] is shown

\includegraphics[scale=0.6]{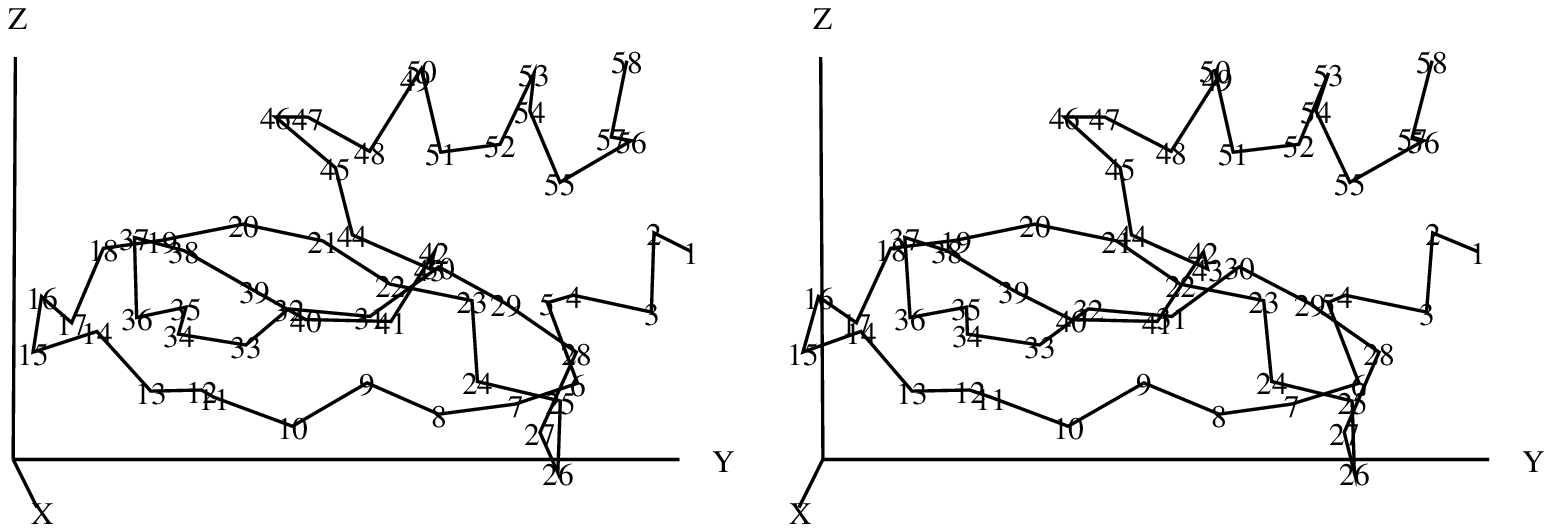}

Figure 1

\footnotesize{Numbered $\alpha$-carbon chain stereoview of the pancreatic trypsin inhibitor protein}.

\normalsize with the atoms positions numbered. The DPSs for this
protein conformation are :

$\{\{(58)(49)(29)(48)(57)(27)(28)(31)(30)(52)\hspace*{2mm}(32)(47)(53)(50)(19)(26)(21)(56)(51)(24)\newline\hspace*{4mm}(33)(20)(23)(55)(46)(25)(22)(34)(54)(18)\hspace*{3.5mm}(1)(45)(17)\hspace*{1.7mm}(5)\hspace*{1.7mm}(6)(35)(44)\hspace*{1.7mm}(2)\hspace*{1.7mm}(8)(43)\newline\hspace*{4mm}(16)\hspace*{1.7mm}(9)(11)\hspace*{1.7mm}(7)\hspace*{1.7mm}(3)(10)(36)\hspace*{1.7mm}(4)(37)(42)\hspace*{2mm}(15)(12)(41)(40)(14)(38)(13)(39)\}_{x},\newline\hspace*{2mm}\{(15)(16)(17)(14)(18)(37)(36)(13)(19)(38)\hspace*{2mm}(34)(35)(12)(11)(39)(20)(33)(46)(10)(40)\newline\hspace*{4mm}(32)(47)(21)(45)(44)\hspace*{1.7mm}(9)(48)(31)(41)(22)\hspace*{2mm}(49)(50)(43)(42)\hspace*{1.7mm}(8)(51)(30)(23)(24)(52)\newline\hspace*{5.9mm}(7)(29)(54)(53)\hspace*{1.7mm}(5)(27)(55)(26)(25)\hspace*{1.7mm}(4)\hspace*{3.6mm}(6)(28)(57)(56)(58)\hspace*{1.7mm}(3)\hspace*{1.7mm}(2)\hspace*{1.7mm}(1)\}_{y},\newline\hspace*{2mm}\{(26)(27)(10)\hspace*{1.7mm}(8)(25)\hspace*{1.7mm}(7)(24)(11)\hspace*{1.7mm}(6)\hspace*{1.7mm}(9)\hspace*{2mm}(12)(28)(13)(33)(34)(15)(31)(32)(29)(17)\newline\hspace*{4mm}(14)(23)(36)(41)(35)\hspace*{1.7mm}(3)(40)\hspace*{1.7mm}(5)(22)(16)\hspace*{2mm}(30)\hspace*{1.7mm}(4)(39)(43)\hspace*{1.7mm}(1)(18)(21)(19)(42)(20)\newline\hspace*{4mm}(44)(38)\hspace*{1.7mm}(2)(37)(55)(48)(45)(51)(52)(57)\hspace*{2mm}(56)(47)(46)(54)(49)(53)(58)(50)\}_{z}\}\hspace*{16mm}(1)$

The main tool developped {[}3-5] in this approach can be described
as a \textbf{fluctuation amplifier}: the small movements of a molecular
system, which are essentailly sampled with the current computer simulating
tools, are encoded by means of a simple combinatorial structure, from
which we can generate the complete set of DPSs corresponding to realizable
3D-conformations that arise from the combination of these movements.
In the preceeding papers {[}3-5] it was described how to build a graph
whose nodes are the cells that are visited by the system in its thermal
wandering, with edges towards the adjacent cells.

As it was it was suggested in {[}1] molecular 3D-conformations are
constrained in a small fraction of the cell volume, for this formalism
to be useful the conformational volume inside a cell has to be probed
and the mean energy of its conformations must be evaluated. The problem
adressed in the present work is how from a DPS code the set of 3D-conformations
it encodes can be reconstructed.

\section{A procedure for embedding molecular conformations in a cubic 3D spatial
lattice}

We start by describing a procedure for embedding the molecular 3D-structures
in a cubic lattice, this can be done using empirical data sampled
from molecular dynamics simulations {[}7]

\subsection*{Procedure 1}

\begin{enumerate}
\item First we determine the dimensions of the lattice by taking as reference
the mean bond length between $C_{\alpha}$ carbons, which is : $3.58$\AA<
$3.86$\AA< $4.13$\AA. In the example developped here we set this
length arbitrarily to $20$ lattice units, which gives a lattice spacing
of $0.19$\AA. 
\item From the range of variation extracted from molecular dynamics any
segment between to lattice points with a length range between $3.58\times20/3.86$
and $4.13\times20/3.86$ is potentially $C_{\alpha}$-$C_{\alpha}$
bond segment. The set of valid lattice bond segments, modulo a lattice
translation along the $x$, $y$ and $z$ axes, is the set of segments
starting at the origin and ending in any lattice point that lies between
two spheres of radius $3.58\times20/3.86$ and $4.13\times20/3.86$
respectively. This gives a total of $1883$ primary segments, i.e.
modulo a reflection through the $xy$, $xz$ and $yz$ planes.
\item Next we determine the range of variation for the bond angles, which
is greater than that for the bond length and varies considerably along
the $C_{\alpha}$chain. For each bond angle $A_{\alpha_{1}\alpha_{2}\alpha_{3}}$along
the main chain we determine two integer numbers : the floored minimum
$\lfloor min(A_{\alpha_{1}\alpha_{2}\alpha_{3}})\rfloor$and the ceiled
maximum range $\lceil max(A_{\alpha_{1}\alpha_{2}\alpha_{3}})\rceil$
respectively. These divide the range $0^{\textdegree}-360^{\textdegree}$
in a number of intervals which in our case give\newline $71\textdegree$-$74\textdegree$-$75\textdegree$-$76\textdegree$-$77\textdegree$-$78\textdegree$-$79\textdegree$-$80\textdegree$-$81\textdegree$-$82\textdegree$-$87\textdegree$-$89\textdegree$-\newline$90\textdegree$-$92\textdegree$-$93\textdegree$-$94\textdegree$-$95\textdegree$-$96\textdegree$-$97\textdegree$-$98\textdegree$-$99\textdegree$-$100\textdegree$-$101\textdegree$-\newline$103\textdegree$-$104\textdegree$-$105\textdegree$-$106\textdegree$-$107\textdegree$-$108\textdegree$-$109\textdegree$-$110\textdegree$-$112\textdegree$-\newline$113\textdegree$-$114\textdegree$-$115\textdegree$-$116\textdegree$-$117\textdegree$-$118\textdegree$-$119\textdegree$-$120\textdegree$-$121\textdegree$-\newline$124\textdegree$-$125\textdegree$-$127\textdegree$-$129\textdegree$-$135\textdegree$-$136\textdegree$-$138\textdegree$-$139\textdegree$-$143\textdegree$-\newline$144\textdegree$-$147\textdegree$-$148\textdegree$-$149\textdegree$-$150\textdegree$-$151\textdegree$-$152\textdegree$-$153\textdegree$-$154\textdegree$-\newline$155\textdegree$-$156\textdegree$-$157\textdegree$-$159\textdegree$-$162\textdegree$-$163\textdegree$-$167\textdegree$\hspace*{24mm}$(2)$

each $A_{\alpha_{1}\alpha_{2}\alpha_{3}}$along the backbone has a
range spanning a given interval set from (2).

\item The set of allowed bond angles formed by pairs of primary segments
are classified according to the following :

\begin{enumerate}
\item Their sign vector : from a broken line formed by two primary segments
with end coordinates $v1=\{x_{0},y_{0},z_{0}\}$ and $v2=\{x_{1},y_{1},z_{1}\}$,
a set of three sign vectors can be generated 

\begin{lyxlist}{00.00.0000}
\item [{$\{sign(-x_{0}),sign(-x_{1}),sign(x_{0}-x_{1})\}$}]~
\item [{$\{sign(-y_{0}),sign(-y_{1}),sign(y_{0}-y_{1})\}$}]~
\item [{$\{sign(-y_{0}),sign(-y_{1}),sign(y_{0}-y_{1})\}$}]~
\end{lyxlist}
where the function $sign(x)$ returns the symbols $+$,$-$ and $0$
if $x$ is positive, negative or zero%
\footnote{The sign matrix and DPSs are equivalent since $sign(x_{0}-x_{1})=+$
means that $x_{0}>x_{1}${[}1].%
}. In order to reduce the size of data we introduce the constraint
that the first three signs of each vector must be $+$, from these
all other sign classes can be generated by reflection symmetry through
the planes perpendicular to $x$, $y$ and $z$. A total of $328$
sign classes are thus generated.

\item Their bond angle interval : pairs of primary bond segments within
a sign class are sorted in angular interval subclasses according to
their bond angle.
\end{enumerate}
\item Next lattice embedded molecular 3D-conformations of the protein backbone
can be generated by joining lattice bond segments with the correct
angular interval and the correct sign matrix between the atoms.
\end{enumerate}

\section{The graph of lattice points}

Each bond between two atoms in the molecular backbone can be approximated
by a lattice primary segment, and each pair of consecutive bonds can
be approximated by a pair of segments having an angle within the bond
angle dynamic range (2). In our approach molecular backbone 3D-conformations
are characterized by the dynamic range of their bond lengths and bond
angles, and by the sign matrix or equivalently the dominance partition
sequence. We have seen in the previous section how to embed the molecular
backbone in a discrete lattice, the problem we try to solve here is
how to enumerate the finite set of lattice 3D-conformations that fulfill
the DPS constraints of a cell or set of cells.

For this we need to build a graphical structure that we call the \textbf{graph of lattice points}\emph{
}in three steps

\subsection*{Procedure 2}

\begin{enumerate}
\item we arbitrarily set the coordinates of the first $C_{\alpha}$ (the
\textbf{root} node) in the backbone to $\{0,0,0\}$ to avoid translation
ambiguities.
\item we choose all segment pairs that have the same sign vectors as the
first $3$ $C_{\alpha}$s and angular value within the interval limits
allowed for the first bond angle.
\item building the $(n+1)^{th}$ lattice bond level in the $C_{\alpha}$
backbone is done from each individual $n^{th}$ level bond segment
by joining it with the second segments of those lattice bond pairs
that : 

\begin{enumerate}
\item the first segment is the same as the $n^{th}$ bond, 
\item the angular value of the pair is within the range of the $(n+1)^{th}$bond
angles.
\end{enumerate}
\item for each level the nodes of the graph are the lattice points at the
upper end of a bond segment. There will be two arcs between any two
points in two consecutive levels if they are connected by a bond segment:
a forward arc from the node in the $n^{th}$ level towards the one
in the $(n+1)^{th}$level if the two are connected by a bond segment,
there is also a reverse arc from the $(n+1)^{th}$level towards the
$n^{th}$level between the same two nodes.
\item The root node has an arc towards every node in the upper level and
reciprocally.
\end{enumerate}

\section{An algorithm for determining the set of inter atomic distances and
weights}

In force fields currently used in molecular dynamics simulations {[}8]
atoms are represented by point-like structures and conformational
energy is calculated with a Hamiltonian which is a sum of two kinds
of terms a) \textbf{local} : which involve groups of $2$, $3$ and
$4$ consecutive atoms needed to calculate the bond, bond angle and
torsion angle energies; b) \textbf{non-local} involving pairs of
atoms lying anywhere in the structure, these are needed for calculating
the electrostatic and van der Waals energy terms.

Usually energy is calculated for only one conformation at a time,
not here : in this work we intend to calculate the energy for a great
number of structures simultaneously, as each term in the hamiltonian
may appear in many structures, \emph{we want to calculate it only
once}. For this we need to know in how many 3D-structures, or equivalently
for how many paths in the graph of lattice points, our term arises.
For the non-local terms we need further to enumerate the whole set
of inter-atomic distances between pairs of atoms, this is possible
on a lattice because squared distances are integer numbers.

We first start by describing an algorithm that assigns two weights
to each node : for a given node $n$ the lower weight ($LW_{n}$)
is the number of downward paths starting at the node and ending at
the root node, the upper weight ($UW_{n}$) is the number of upward
paths starting at the node and ending at some node in the top level.
The algorithm is as follows :

\subsection*{Procedure 3}

\begin{enumerate}
\item For the root node we set its lower weight value to $1$.
\item We go to the next level.
\item For each node $n$ in the level we set the $LW_{n}$ value to the
sum of its downward link nodes $LW$s.
\item If the top level has not been reached we go to step 2. 
\item Otherwise for every node in the top level we set the upper weight
to $1$.
\item We go to the previous level.
\item For each node $n$ in the level we set the $UW_{n}$ value to the
sum of its upward link nodes $UW$s.
\item If the root level has not been reached we go step 6.
\item Otherwise we terminate the procedure.
\end{enumerate}
Thus two consecutive lattice points $a$ and $b$ will contribute
the quantity $E_{bond}(d_{a,b})\times LW{}_{a}\times UW_{b}$ to the
global bond energy, where $d_{a,b}$is the length of the lattice segment
between $a$ and $b$ and $LW_{a}$ and $UW_{b}$ are the lower and
upper weights of $a$ and $b$ respectively. Similarly for the bond
and torsion energy of consecutive lattice points $a$, $b$, $c$
and $d$ we have $E_{angle}(\theta_{a,b,c})\times LW_{a}\times UW_{c}$
and $E_{torsion}(\phi_{a,b,c,d})\times LW_{a}\times UW_{d}$ respectively.
These quantities can be calculated using a variant of the basic algorithm
described in procedure 1.

To determine the weights for distances between arbitrary pairs of
nodes we need three new data structures : an integer flag variable
in every node, a sorted table of distances and a register array whose
lengh is the number of levels.

\subsubsection*{Procedure 4}

\begin{enumerate}
\item We set the flag in every node to $0$.
\item For the $N_{top}$ nodes in the top level.
\item For every node in the top level

\begin{enumerate}
\item We assign to each one a number $n_{top}$ from $1$ to $N_{top}$.
\item We enter the node in the top level of the array register.
\item We follow every downward link in succesion to the previous level.
\end{enumerate}
\item For each node thus reached 

\begin{enumerate}
\item We set its flag to the top level number and we enter the node in the
corresponding level of the array register.
\item We follow every downward link in succesion.
\item Upon reaching the root level or a node whose flag value is equal to
the current $n_{top}$

\begin{enumerate}
\item We calculate the distances between all nodes in the array register
from the current to the top level.
\item Every calculated distance $d_{a,b}$ between nodes $a$ and $b$ in
levels $L_{a}>L_{b}$ is assigned an upper and lower weight : $UW_{a}$
and $LW_{b}$. The array $\{d_{a,b},LW_{b},UW_{a}\}$ is searched
in the table of distances

\begin{enumerate}
\item If not found it is entered in the table together with the number $count_{a,b}$
which is set to $1$.
\item If the array already exists in the table $count_{a,b}$ is increased
by $1$.
\end{enumerate}
\item We return to the previous node and from there we follow the next downward
link to a node in the lower level.
\end{enumerate}
\item Upon reaching a node in level $L_{v}$ whose flag value $n>0$ is
different from the current $n_{top}$

\begin{enumerate}
\item We continue the downward exploration of the graph setting the flag
of every node to $n_{top}$.
\item We upon reaching the root node we do not compute distances between
nodes in levels $L_{v}$ or lower.
\end{enumerate}
\end{enumerate}
\item The procedure terminates when all downward links in every top node
have been explored.
\end{enumerate}
Each distance $d_{a,b}$ from the table of distances will have a weigth
$W_{a,b}=UW_{a}+LW_{b}+count_{a,b}$ which will we used to compute
the van der Waals and electrostatic terms of the hamiltonian $E_{vdW}(d_{a,b})\times W_{a,b}$
and $E_{elec}(d_{a,b})\times W_{a,b}$ respectively.

\section{Discussion}

The procedures discussed in this work make possible removal of two
important hurdles of the formalism in its way towards practical applications
: 

\begin{enumerate}
\item The construction of realistic 3D-conformations with a given partition
sequence. As was discussed in {[}1], a lot of structural information
disappears when replacing the coordinates of a molecule by the inverse
sequences of its DPSs, however though the molecule appears heavily
deformed all the secondary structure motifs : $\alpha$-helices, $\beta$-sheets,
turns, ... together with the overall 3D folding arrangement can still
be recognized. This means that the $3\times(N-1)$-dimensional volume%
\footnote{For a molecule with $N$ atoms.%
} of a cell in conformational space is big enough to allow for very
lean codes, but it still exerts a constraint that keeps the 3D-structures
close to the real ones. 
\item The combinatorial structures described above for building molecular
conformations can be transformed to calculate their mean energy. Which
is essential for comparing the results from this formalism with experimental
data.
\end{enumerate}
It is also quite conceivable that in some practical situations the
procedures decribed above may overwhelmed by the amount of data generated,
in this case the combinatorial structures they generate will have
to be pruned in order to make them useful. This question will be adressesed
when submitting this model of conformational molecular space to phenomenological
tests.

\end{document}